\documentclass[preprint]{elsarticle}
\usepackage[margin=1in]{geometry}
\usepackage{dsfont}
\usepackage{amsmath,amsfonts,amssymb}
\usepackage{graphicx}
\usepackage[font=footnotesize]{caption}
\usepackage{subcaption}
\usepackage{hyperref}
\newcommand{\PN}{\ensuremath{P_{N}}}
\newcommand{\SN}{\ensuremath{S_{N}}}
\newcommand{\grad}{\ensuremath{\vec{\nabla}}}
\newcommand{\vo}{\ensuremath{\vec{\Omega}}}
\newcommand{\vr}{\ensuremath{\vec{r}}}
\newcommand{\vn}{\ensuremath{\vec{n}}}
\newcommand{\pD}{\ensuremath{\partial\mathcal{D}}}
\newcommand{\D}{\ensuremath{\mathcal{D}}}
\newcommand{\drm}{\ensuremath{\mathrm{d}}}
\newcommand{\Sp}{\ensuremath{\mathcal{S}}}
\newcommand{\sigt}{\ensuremath{\sigma_{t}}}
\usepackage{alphalph}

\usepackage{epstopdf}
\usepackage{stmaryrd}
\usepackage{float}
\usepackage{color}
\usepackage{cases}
\usepackage{lineno}

\newcommand{\revision}[1]{\textcolor{black}{#1}}

\hypersetup{
	colorlinks,
	citecolor=black,
	filecolor=black,
	linkcolor=blue,
	urlcolor=black
}
\DeclareGraphicsExtensions{.pdf,.png,.jpg,.eps,.bmp}

\begin{document}
	
\begin{frontmatter}
\title{Globally Conservative, Hybrid Self-Adjoint Angular Flux and Least-Squares Method Compatible with Void \tnoteref{nsf-support}}
\date{\today}
\author[tamu-ne]{Vincent M. Laboure\corref{cor}}
\ead{vincent.laboure@tamu.edu}

\author[tamu-ne]{Ryan G.\,McClarren}
\ead{rgm@tamu.edu}

\author[inl]{Yaqi Wang}
\ead{yaqi.wang@inl.gov}

\cortext[cor]{Corresponding author. Tel.:+1 979 224 8506}

\tnotetext[nsf-support]{This material is based, in part, upon work supported by the National Science Foundation under Grant No. 1217170.  The research of the third author is sponsored by the U.S. Department of Energy, under DOE Idaho Operations Office Contract DE-AC07-05ID14517.}

\address[tamu-ne]{Nuclear Engineering Department,
	Texas A\&M University
	College Station, TX 77843}

\address[inl]{Reactor Physics and Analysis \\
	Idaho National Laboratory, Idaho Falls, Idaho, USA}

\begin{abstract}
	In this paper, we derive a method for the second-order form of the transport equation that is both globally conservative and compatible with voids, using Continuous Finite Element Methods (CFEM).
	The main idea is to use the Least-Squares (LS) form of the transport equation 
	%compatible with voids \cite{ls-sn} 
	in the void regions and the Self-Adjoint Angular Flux (SAAF) form %\cite{Morel1999,JimE.Morel2006,Liscum-Powell2000,Liscum-Powell2002} 
	elsewhere.
	While the SAAF formulation is globally conservative, the LS formulation need a correction in void. The price to pay for this fix is the loss of symmetry of the bilinear form.
	We first derive this Conservative LS (CLS) formulation in void. Second we combine the SAAF and CLS forms and end up with an hybrid SAAF--CLS method, having the desired properties.
	We show that extending the theory to near-void regions is a minor complication and can be done without affecting the global conservation of the scheme. Being angular discretization agnostic, this method can be applied to both discrete ordinates (\SN) and spherical harmonics (\PN) methods.
	However, since a globally conservative and void compatible second-order form already exists for \SN\,(Wang et al.~2014), 
	but is believed to be new for \PN, we focus most of our attention on that latter angular discretization.
	We implement and test our method in Rattlesnake within the Multiphysics Object Oriented Simulation Environment (MOOSE) framework. 
	%\cite{MOOSE}
	Results comparing it to other methods are presented. 
\end{abstract}

% \textbf{Keywords}:
	
\end{frontmatter}

%\maketitle
%\tableofcontents
%\linenumbers

\section{Introduction}
\label{sect::intro}

Second-order forms of the linear Boltzmann transport equation typically perform well with neutronics calculations and can be preferred over first-order forms for several reasons, among which we wish to emphasize two.
First, they directly allow for the use of CFEM \cite{Morel1999}, whereas first-order forms require some sort of stabilization, often added using Discontinuous Finite Element Methods (DFEM) and an appropriate numerical flux \cite{reed1973triangularmesh}.
While this latter discretization can give very satisfying results even for non-smooth solutions, in particular in convection dominated regimes \cite{DG2001, GuermondKanschat2010, LarsenMorelMiller2001, AdamsDFE, LaboureMcClarrenHauck2016}, its comparatively higher cost often makes it unattractive if the solution is smooth enough.
Second, they may result in a symmetric positive definite (SPD) matrix, thus suited for the use of conjugate-gradient (CG) Krylov solvers \cite{Morel1999, GolubVanLoan1996}.

Nevertheless, second-order forms are not exempt from any flaws. 
Most of them, such as the Self-Adjoint Angular Flux (SAAF) \cite{Morel1999,JimE.Morel2006,Liscum-Powell2000,Liscum-Powell2002,WangNDA} or the even-parity \cite{LewisMiller1984} formulations, formally break down in void regions because it requires the evaluation of $\sigt^{-1}$, the inverse of the total cross-section. 
Approximating a zero cross-section with an arbitrary low number is not a viable solution as it drastically degrades the solver performance \cite{Drumm2011}. 
An alternative is the SAAF formulation with a Void Treatment (SAAF--VT) \cite{WangNDA}, one of the downsides of this method being that it no longer results in an SPD system. 
Furthermore, it is not well suited to a \PN\,expansion because the bilinear form in the void regions goes -- in steady state and as the mesh is refined -- to the first-order form in void, which is ill-conditioned for \PN\,\cite{LaboureMcClarrenHauck2016}.

Another class of methods based on weighted residual minimizations, such as Least-Squares (LS) formulations, may not have this formal problem when $\sigma_t$ goes to zero but are in general not globally conservative. 
This is in particular the case of the LS method compatible with voids \cite{ls-sn} and can significantly degrade the accuracy of the solution, in particular for $k$-eigenvalue problems \cite{LaboureWang2016}. 
Acceleration schemes such as Nonlinear Diffusion Acceleration (NDA) can help recover global conservation \cite{WangNDA} but such schemes have yet to be developed for \PN\,methods.

Therefore, there does not exist -- to our knowledge -- any second-order form for \PN\,that would result in a globally conservative and void compatible scheme. 
The driving purpose of the present work is to develop such a method, although this work will not be limited to \PN. The idea is to decompose our domain into two regions: a non-void region noted $\D_1$ discretized using the standard SAAF formulation, which is known to be globally conservative; and a void region noted $\D_0$ using the LS formulation compatible with voids. 

In this paper, we show that, unfortunately, the LS formulation in void is \revision{not globally conservative because of the discretization error in $\D_0$.}
A conservative fix is derived, yielding the Conservative LS (CLS) method. 
It  can also be extended to treating near-void regions (with uniform cross-section).
However, this additional term -- just like the void treatment of the SAAF--VT method \cite{WangNDA} -- breaks the symmetry of the bilinear form. 
That being done, the variational formulation of the hybrid SAAF--CLS scheme is derived and is shown to be globally conservative.

The remainder of this paper is structured as follows. 
In Section \ref{sec:SAAFCLS}, we show why the LS formulation is not globally conservative and how to make it so. We then derive the SAAF--CLS method, achieving void compatibility and global conservation with an appropriate choice of the scaling between the SAAF and CLS terms. 
In Section \ref{sec:nearVoid}, we no longer require $\sigt$ to be zero in the void regions but only to be constant therein,  particularly addressing the treatment of near-void regions.
In Section \ref{sec:results}, we discuss the actual implementation of the method and study the results on: 
(i) a slab geometry pure absorber problem, specifically looking at the importance of the conservative fix; 
(ii) a multigroup heterogeneous $k$-eigenvalue problem with a void region, comparing the SAAF--CLS--\PN\,and SAAF--CLS--\SN\,methods to reference solutions as well as to already existing void-compatible methods, such as LS--\PN, LS--\SN\,and SAAF--VT--\SN; and 
(iii) a near-void benchmark problem introduced by Kobayashi et al \cite{Kobayashi2001}. Section \ref{sec:conclusion} presents the conclusions of this work and suggests some future studies.

%The \PN\,method can be seen as a spectral Galerkin Finite Element method using the spherical harmonics as both the test and basis functions.

%Next, we choose to use a truncated spherical harmonics (\PN) expansion to approximate the angular dependency of the solution. This ensures in particular that the numerical scalar flux is immune to ray-effects but -- among other flaws -- it can potentially become oscillatory and/or negative \cite{Brunner2002, McClarrenHauck2010, HauckMcClarren2010, Radice2013}. The odd moments can be simply expressed as a function of the even moments under the condition that $\sigma_t$ is non-zero. The number of unknowns can then be reduced almost by half but may induce a loss of accuracy.

\section{Void Compatible SAAF--CLS Method}
\label{sec:SAAFCLS}

\subsection{Problem and notation}
We use the following notation: considering a spatial domain $\mathcal{V}$ with boundary $\partial\mathcal{V}$, we define the following operators:
\begin{equation*}
\left(a,b\right)_{\mathcal{V}} \equiv \int_{\mathcal{V}}\int_{\Sp} a(\vr,\vo)b(\vr,\vo)\; \drm\Omega \drm r \quad,\quad \langle a,b\rangle^+_{\partial\mathcal{V}} \equiv \int_{\partial\mathcal{V}} \int_{\vo\cdot\vn(\vr)>0}  a(\vr,\vo)b(\vr,\vo)\; |\vo\cdot\vn|\drm\Omega \drm r,
\end{equation*}
\begin{equation*}
\langle a,b\rangle_{\partial\mathcal{V}} \equiv \int_{\partial\mathcal{V}} \int_{\Sp}  a(\vr,\vo)b(\vr,\vo)\; \vo\cdot\vn\,\drm\Omega \drm r \;,\;
\langle a,b\rangle^-_{\partial\mathcal{V}} \equiv \int_{\partial\mathcal{V}} \int_{\vo\cdot\vn(\vr)<0}  a(\vr,\vo)b(\vr,\vo)\; |\vo\cdot\vn|\drm\Omega \drm r,
\end{equation*}
with $\vec{r}$ and $\vec{\Omega}$ being, respectively, the spatial and angular coordinates and $\vn$ being the outward unit normal vector to $\partial\mathcal{V}$ and $\Sp$ being $\{\mu\in[-1,1]\}$, $\{(\mu,\varphi)\in[-1,1]\times[0,\pi)\}$  and $\{(\mu,\varphi)\in[-1,1]\times[0,2\pi)\}$ if the problem depends on one, two and three spatial dimensions respectively. 
%\vml{Is there a problem with that previous sentence??}
We also define $w=\int_{\Sp}\drm\Omega$ where $\drm\Omega = \drm\mu\drm\varphi$. 

For the derivation, we are considering -- for simplicity\footnote{Extending the present theory to the multigroup transport equation does not pose any problems.} % Extending it time-dependent problems might be a little more tricky, unless it can be assumed that the speed of the particles in void is fast enough, so that it can be considered that any particle entering a void region exits immediately at the other end of the void region.} 
-- the following one-group steady-state transport problem:
\begin{equation}
\vec{\Omega}\cdot\vec{\nabla}\Psi + \sigt(\vec{r})\Psi(\vr,\vo) = \int_{\Sp} \sigma_s (\vr,\vo^\prime\rightarrow\vo) \Psi (\vr,\vo^\prime) \, \mathrm{d}\Omega^\prime +  \nu\sigma_f(\vr) \Phi(\vr) + S(\vr,\vo),
\label{eq0}
\end{equation}
where $\Psi$ and $\Phi$ represent respectively the angular and scalar flux. 
In addition, $\sigma_t$, $\sigma_s$ and $\sigma_f$ respectively denote the total, scattering and fission macroscopic cross-sections and $\nu$ is the average number of neutrons emitted per fission. $S$ is the volumetric source. 
Boundary conditions are specified at the boundary of the domain \pD~for incoming directions, i.e.~$\Psi(\vr_b,\vo)  \equiv \Psi^{\text{inc}}(\vr_b,\vo)$ for $\vr_b\in\pD$ and $\vo\cdot\vn(\vr)<0$.
We can rewrite Eq.\,\eqref{eq0} in operator form:
\begin{equation}
L\Psi =  H\Psi + S,
\label{eq1}
\end{equation}
where $L$ is the streaming plus collision operator and $H$ is the fission plus scattering operator.\\\\
We decompose the spatial domain as $\D = \D_1 \uplus \mathcal{D}_0$ where $\D_0$ is such that $\sigt = 0$. 
The interface between $\D_0$ and $\D_1$ is noted $\Gamma = \D_0\cap\D_1$. 
We refer to the continuous finite element space corresponding to $\D$, $\D_0$ and $\D_1$ as $V$, $V_0$ and $V_1$, respectively. 
Besides, $\Psi^\star$ designates a test function.

\subsection{Non-conservativity of the LS method}
We start with the LS formulation in void applied to $\D_0$ \cite{ls-sn, PetersonHammer2015}: find $\Psi\in V_0$ such that for all $\Psi^*\in V_0$,
\begin{equation}
\left(\vo\cdot\grad\Psi^\star, \vo\cdot\grad\Psi \right)_{\D_0} 
+ \langle c \Psi^\star, (\Psi-\Psi^{\text{inc}}) \rangle_{\partial\mathcal{D}_0}^{-} = 0,
\end{equation}
where $c>0$ is a constant\footnote{Note that the LS method does not fundamentally require $c$ to be constant but the following reasoning does.} used to weakly impose the boundary conditions, with units of a cross-section. 
Using the divergence theorem to transform $\langle c \Psi^\star, \Psi \rangle_{\partial\mathcal{D}_0}^{-}$, it becomes:
\begin{equation}
\left(\vo\cdot\grad\Psi^\star, \vo\cdot\grad\Psi \right)_{\D_0}  - \left(c \Psi^\star, \vo\cdot\grad\Psi \right)_{\D_0} - \left(\vo\cdot\grad\Psi^\star, c\Psi \right)_{\D_0}
+  \langle c \Psi^\star, \Psi \rangle_{\partial\mathcal{D}_0}^{+} - \langle c \Psi^\star, \Psi^{\text{inc}} \rangle_{\partial\mathcal{D}_0}^{-} = 0.
\end{equation}
In particular, for the constant test function $\Psi^\star=1$, we have:
\begin{equation}
- \left(c , \vo\cdot\grad\Psi \right)_{\D_0} +   \langle c , \Psi \rangle_{\partial\mathcal{D}_0}^{+} - \langle c, \Psi^{\text{inc}} \rangle_{\partial\mathcal{D}_0}^{-} = 0,
\label{eq:LSconservartion}
\end{equation}
\revision{which exhibits the lack of global conservation of the scheme. Indeed, conservation in $\D_0$ is achieved if and only if $(1 , \vo\cdot\grad\Psi )_{\D_0} = 0$, i.e.\,if the discretization error in $\D_0$ is negligible.} 
While the analytical solution does satisfy this relation, nothing can be said about the numerical solution. 
\revision{Thus, this LS formulation constitutes a consistent, yet not globally conservative discretization in $\D_0$.}

\subsection{Conservative Least-Squares method}
From the previous expression, it is clear that the scheme would be globally conservative if we were to add $(c \Psi^\star, \vo\cdot\grad\Psi )_{\D_0}$ to the variational formulation. 
Since this term can be obtained directly from Eq.\,\eqref{eq1} in $\D_0$, the converged solution would not be affected by this change. 
We therefore define the CLS formulation applied to $\D_0$ to be: find $\Psi\in V_0$ such that for all $\Psi^*\in V_0$,
\begin{equation} 
\left(\vo\cdot\grad\Psi^\star, \vo\cdot\grad\Psi \right)_{\D_0} + \left(c \Psi^\star, \vo\cdot\grad\Psi \right)_{\D_0}
+ \langle c \Psi^\star, (\Psi-\Psi^{\text{inc}}) \rangle_{\pD_0}^{-} = 0.
\label{eq:conservativeFix}
\end{equation}
Alternatively, splitting the boundary terms depending on whether they belong to $\pD$ or to $\Gamma$, it can be expressed as: 
\begin{equation} \label{CLS}
\left(\vo\cdot\grad\Psi^\star, \vo\cdot\grad\Psi \right)_{\D_0} - \left(\vo\cdot\grad\Psi^\star, c \Psi \right)_{\D_0}
+ \langle c \Psi^\star, \Psi \rangle_{\pD^0}^{+} - \langle c \Psi^\star, \Psi^{\text{inc}} \rangle_{\pD^0}^{-} + \langle c \Psi^\star, \Psi \rangle_{\Gamma}^{+,0} - \langle c \Psi^\star, \Psi^{\text{inc}} \rangle_{\Gamma}^{-,0} = 0,
\end{equation}
where $\pD^0 \equiv \pD\cap\pD_0$. 
In this expression, we have also used the notation $\langle\cdot,\cdot\rangle^{\pm,0}_{\Gamma}$ to indicate that the angular integration half-range $\pm\vo\cdot\vn(\vr)>0$ is determined with $\vn$ being the outward unit vector normal to $\Gamma$ with respect to $\D_0$ (i.e.\,locally pointing towards $\D_1$). 
This formulation is globally conservative but is not symmetric. 

\subsection{SAAF--CLS method}
Choosing $\Psi = \sigt^{-1}(-\vo\cdot\grad\Psi+H\Psi+S)$ as the angular flux equation (the so-called first AFE in \cite{WangNDA}), the SAAF formulation applied to $\D_1$ is given by: find $\Psi\in V_1$ such that for all $\Psi^*\in V_1$,
\begin{equation} \label{SAAF}
\begin{split}
&\left(\vo\cdot\grad \Psi^\star,\dfrac{1}{\sigt}\,\vo\cdot\grad \Psi\right)_{\D_1} 
+ \left(\sigma_t \Psi^\star,\Psi\right)_{\D_1}\\
&+\langle\Psi^\star,\Psi\rangle^+_{\pD^1}  - \langle\Psi^\star,\Psi^{\text{inc}}\rangle^-_{\pD^1} + \langle\Psi^\star,\Psi\rangle^{+,1}_{\Gamma}  - \langle\Psi^\star,\Psi^{\text{inc}}\rangle^{-,1}_{\Gamma}= \left(\dfrac{1}{\sigt}\,\vo\cdot\grad \Psi^\star + \Psi^\star,H\Psi + S\right)_{\D_1},
\end{split}
\end{equation}
where $\pD^1 \equiv \pD\cap\pD_1$. 
We scale Eq.\,\eqref{CLS} with a constant \revision{$\xi>0$} with units of a cross-section, for consistency.
We then combine it with Eq.\,\eqref{SAAF} and notice that $\Psi$ is continuous across $\Gamma$ (i.e.\,$\Psi=\Psi^{\text{inc}}$ on $\Gamma$) to end up with:
\begin{equation}
\begin{split}
&\left(\vo\cdot\grad \Psi^\star,\dfrac{1}{\sigt}\,\vo\cdot\grad \Psi\right)_{\D_1}
+ \left(\sigt \Psi^\star,\Psi\right)_{\D_1}  + \left(\vo\cdot\grad\Psi^\star, \dfrac{1}{\xi}\,\vo\cdot\grad\Psi \right)_{\D_0} - \left(\vo\cdot\grad\Psi^\star, \dfrac{c}{\xi} \Psi \right)_{\D_0}\\
&+\langle \dfrac{c}{\xi} \Psi^\star, \Psi \rangle_{\pD^0}^{+} - \langle \dfrac{c}{\xi} \Psi^\star, \Psi^{\text{inc}} \rangle_{\pD^0}^{-} + \langle \dfrac{c}{\xi} \Psi^\star, \Psi \rangle_{\Gamma}^{0} \\
&+\langle\Psi^\star,\Psi\rangle^+_{\pD^1}  - \langle\Psi^\star,\Psi^{\text{inc}}\rangle^-_{\pD^1} + \langle\Psi^\star,\Psi\rangle^{1}_{\Gamma} = \left(\dfrac{1}{\sigt}\,\vo\cdot\grad \Psi^\star + \Psi^\star,H\Psi + S\right)_{\D_1}.
\end{split}
\end{equation}
Global conservation imposes the following condition:
\begin{equation}
c = \xi.
\end{equation}
The SAAF--CLS weak formulation is then given by: find $\Psi\in V$ such that for all $\Psi^*\in V$,
\begin{equation} \label{SAAFCLS}
\begin{split}
&\left(\vo\cdot\grad \Psi^\star,\dfrac{1}{\sigt}\,\vo\cdot\grad \Psi\right)_{\D_1}
+ \left(\vo\cdot\grad\Psi^\star, \dfrac{1}{c}\,\vo\cdot\grad\Psi \right)_{\D_0} + \left(\sigma_t \Psi^\star,\Psi\right)_{\D_1} - \left(\vo\cdot\grad\Psi^\star, \Psi \right)_{\D_0}\\
&+\langle \Psi^\star, \Psi \rangle_{\pD}^{+} - \langle \Psi^\star, \Psi^{\text{inc}} \rangle_{\pD}^{-}  = \left(\dfrac{1}{\sigt}\,\vo\cdot\grad \Psi^\star + \Psi^\star,H\Psi + S\right)_{\D_1}.
\end{split}
\end{equation}
One can check that this scheme is globally conservative by choosing $\Psi^\star=1$. 
Interestingly, the non-symmetric term is very similar to the extra term in the SAAF--VT formulation \cite{WangNDA}. 
The difference between the two formulations however is that in void regions, the second-order streaming term over $\D_0$ in Eq.\,\eqref{SAAFCLS} do not vanish, even when the mesh is infinitely refined. 
This is crucial to avoid having a singular term when using a \PN\,expansion.

\subsection{Value of $c$}
\label{sec:valueOfc}

Numerically, changing the value of $c$ can have a very significant impact on the solver convergence. 
While the optimal value for $c$ is still an open question, it is interesting to note that it has the same units as a cross-section. 
It is therefore wise to choose it in the order of $\sigma_t$ in the non-void regions. 
In this paper, it is chosen to be $c=1$ cm$^{-1}$, unless otherwise specified.

\revision{We further explore this issue in Section \ref{sec:ReedsProblem} and show that even if $\sigma_t$ strongly varies on $\Gamma$, the choice for $c$ has a small effect on the solution, especially outside the void region.}

\section{Extension to Uniform Non-Void Regions}
\label{sec:nearVoid}
In this section, we show that we can similarly derive a void compatible, globally conservative scheme in the more general setting of a uniform non-void region in $\D_0$. 
The importance of this result lies in the fact that real-world applications rarely contain pure void regions but more realistically near-void regions. 
The only different assumption is thus that we no longer require $\sigt = 0$ in $\D_0$ but only to be uniform therein, i.e.\,$\sigt = \sigma_{0}$ in $\D_0$. 
Although the driving application is the treatment of near-void regions, we do not need to assume that $\sigma_{0}$ is small for the reasoning in this section to hold. 

\subsection{Generalized CLS method}
The LS formulation compatible with voids applied to $\D_0$ is now given by \cite{ls-sn, PetersonHammer2015}:
\begin{equation} \label{LS}
\left(L\Psi^\star, L\Psi \right)_{\D_0}
+ \langle c \Psi^\star, (\Psi-\Psi^{\text{inc}}) \rangle_{\pD_0}^{-} = \left(L\Psi^\star, H\psi + S \right)_{\D_0}.
\end{equation}
It can be shown that the standard SAAF formulation and Eq.\,\eqref{LS} are equivalent if and only if $\sigt$ is a strictly positive constant and if we have $c=\sigt$. 
Under these two conditions, Eq.\,\eqref{LS} is globally conservative. 
While the first condition is included in our assumptions, we cannot satisfy the second in all generality because the boundary terms would vanish in void or near-void regions. 
If $c\neq\sigt$, it is possible to show consistency between Eq.\,\eqref{LS} and a SAAF-like globally conservative formulation, in the sense that both schemes would be equivalent as the discretization error goes to zero. 
In particular, the choice $c =\max\left(\sigma_t,\varsigma/h\right)$, where $h$ characterizes the mesh size and $\varsigma$ is a constant, typically chosen to be 2, is consistent with the SAAF--VT formulation \cite{WangNDA}, as shown in \cite{LaboureWang2016}. 
In summary, this means that, unless $c=\sigt$, Eq.\,\eqref{LS} \revision{constitutes a consistent, yet not globally conservative discretization, even though $\sigt$ is constant over $\D_0$.}

Just as in Section \ref{sec:SAAFCLS}, we can define the Conservative Least-Squares formulation on $\D_0$ by adding the term $((c-\sigma_{0})\Psi^*, \vo\cdot\grad\Psi + \sigma_{0}\Psi-H\psi-S )_{\D_0}$ to the LS formulation, which is consistent with the transport equation. 
After some manipulations, the formulation can be expressed as: find $\Psi\in V_0$ such that for all $\Psi^*\in V_0$,
\begin{equation}\label{CLSnearvoid}
\begin{split}
&\left(\vo\cdot\grad\Psi^\star,\vo\cdot\grad\Psi \right)_{\D_0} + \left(c\Psi^\star,\sigma_{0}\Psi \right)_{\D_0}
+ \langle c \Psi^\star, \Psi \rangle_{\pD_0}^{+} - \langle c \Psi^\star, \Psi^{\text{inc}} \rangle_{\pD_0}^{-} \\
&- \left((c-\sigma_{0})\vo\cdot\grad\Psi^\star, \Psi  \right)_{\D_0}  = (\vo\cdot\grad\Psi^\star + c\Psi^\star,H\Psi+S)_{\D_0}.
\end{split}
\end{equation}
One can check that this scheme is globally conservative by choosing $\Psi^\star=1$ and that it does reduce to Eq.\,\eqref{CLS} in void.

\subsection{Generalized SAAF--CLS method}
The generalized SAAF--CLS formulation is similarly obtained by scaling Eq.\,\eqref{CLSnearvoid} with $1/\xi$ and adding it to Eq.\,\eqref{SAAF}. 
Global conservation requires the same condition as before, namely $c=\xi$. 
The weak form is then given by: find $\Psi\in V$ such that for all $\Psi^*\in V$,
\begin{equation} \label{SAAFCLSnearvoid}
\begin{split}
&\left(\vo\cdot\grad \Psi^\star,\dfrac{1}{\sigt}\,\vo\cdot\grad \Psi\right)_{\D_1} + \left(\vo\cdot\grad\Psi^\star, \dfrac{1}{c}\,\vo\cdot\grad\Psi \right)_{\D_0} + \left(\sigma_t \Psi^\star,\Psi\right)_{\D_1} + \left(\Psi^\star,\sigma_0\Psi \right)_{\D_0}   \\
& - \left(\left(1-\dfrac{\sigma_{0}}{c}\right)\vo\cdot\grad\Psi^\star, \Psi  \right)_{\D_0} +\langle\Psi^\star,\Psi\rangle^+_{\pD}  - \langle\Psi^\star,\Psi^{\text{inc}}\rangle^-_{\pD} \\
&\quad = \left(\dfrac{1}{\sigt}\,\vo\cdot\grad \Psi^\star + \Psi^\star,H\Psi + S\right)_{\D_1} + \left(\dfrac{1}{c}\,\vo\cdot\grad \Psi^\star + \Psi^\star,H\Psi+S\right)_{\D_0}.
\end{split}
\end{equation}
Once again, this formulation is globally conservative, non-symmetric and reduces to Eq.\,\eqref{SAAFCLS} if $\D_0$ is pure void. It is interesting to note that this formulation is identical to the SAAF-VT formulation \cite{WangNDA} with $\tau = 1/\sigt$ in $\D_1$ and $\tau = 1/c$ in $\D_0$ (see Eq.~\eqref{eq:SAAFVT}).

\section{Numerical Results}
\label{sec:results}

\subsection{Implementation}
The method derived above is implemented in Rattlesnake, the transport solver from the Idaho National Laboratory based on the MOOSE framework \cite{MOOSE}. 
All the results presented below are obtained with the first order LAGRANGE elements from libMesh \cite{libMeshPaper}. 
We use the Preconditioned Jacobian Free Newton Krylov (PJFNK) method for the nonlinear solves with the PETSc \cite{PETSc} restarted generalized minimal residual (GMRES) solver for the linear solves, \revision{with the restart parameter set to 100.} 
Preconditioning is done through the algebraic multigrid Hypre BoomerAMG \cite{hypre} preconditioner. \revision{In a few instances, as mentioned in Section \ref{sec:kobayashi},
% for the SAAF--CLS--\PN~method and large values of $N$),
the built-in block Jacobi preconditioner in PETSc  appeared to be more efficient.}\footnote{\revision{Note however that the iteration counts shown in Table \ref{tab::VTvsCLS} are all obtained with the same preconditioner (Hypre BoomerAMG).}}

\revision{The eigenvalue problems are solved using the Inverse Power Method with Chebyshev acceleration with the same solver and preconditioner as previously mentioned.}

\subsection{Terminology}
In the following sections, several methods are being compared. Here, we specify what is precisely meant by each of these. SAAF--CLS refers to Eq.\,\eqref{SAAFCLSnearvoid} (or, equivalently, Eq.\,\eqref{SAAFCLS} if $\sigma_0 = 0$).
To highlight why the conservative fix introduced in Eq.\,\eqref{eq:LSconservartion} is crucial, we also show the results of the same method without the conservative fix and refer to it as SAAF--LS. 
For convenience, we recall the SAAF--VT formulation \cite{WangNDA}: find $\Psi\in V$ such that for all $\Psi^*\in V$,
\begin{equation}
\begin{split}
\left(\vo\cdot\grad \Psi^\star,\tau\,\vo\cdot\grad \Psi\right)_{\D} 
&- \left(\vo\cdot\grad \Psi^\star,(1-\tau\sigma_t)\Psi\right)_{\D} 
+ \left(\sigma_t \Psi^\star,\Psi\right)_{\D} \\
&+\langle\Psi^\star,\Psi\rangle^+_{\pD}  - \langle\Psi^\star,\Psi^{\text{inc}}\rangle^-_{\pD}= \left(\tau\,\vo\cdot\grad \Psi^\star + \Psi^\star,H\Psi + S\right)_{\D},
\end{split}
\label{eq:SAAFVT}
\end{equation}
with $\tau$ being defined as $\tau = \min\left(\sigma_t^{-1},h/\varsigma\right)$, where $h$ characterizes the mesh size and $\varsigma$ is a constant, typically chosen to be 2.

Lastly, we consider the plain LS method which is obtained using Eq.\,\eqref{LS} over the whole domain. By default, we choose $c=1/\tau$, which is consistent with the SAAF--VT formulation in the case of a constant $\tau$ \cite{LaboureWang2016}.

\subsection{Slab geometry pure absorber problem}
\label{sec:1Dproblem}

\begin{figure}
	%\centering
	\begin{subfigure}[b]{0.49\textwidth}
		%\hspace*{-2cm}
		\includegraphics[width=\textwidth]{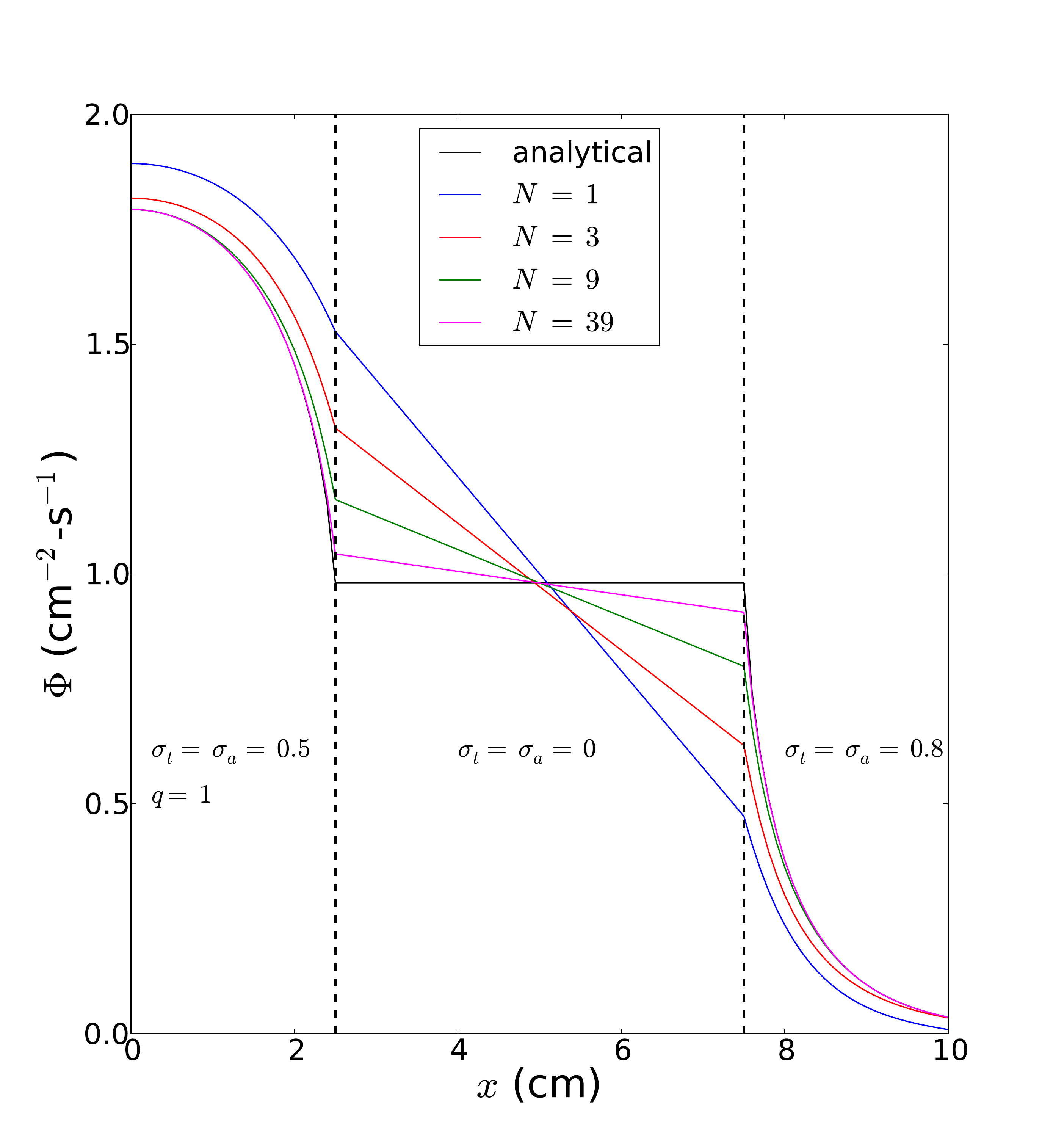}
		\subcaption{SAAF--LS--\PN}
		\label{fig:1Dpb_LS}
	\end{subfigure}%
	%\quad %add desired spacing between images, e. g. ~, \quad, \qquad, \hfill etc.
	%(or a blank line to force the subfigure onto a new line)
	\begin{subfigure}[b]{0.49\textwidth}
		%\hspace*{-2cm}
		\includegraphics[width=\textwidth]{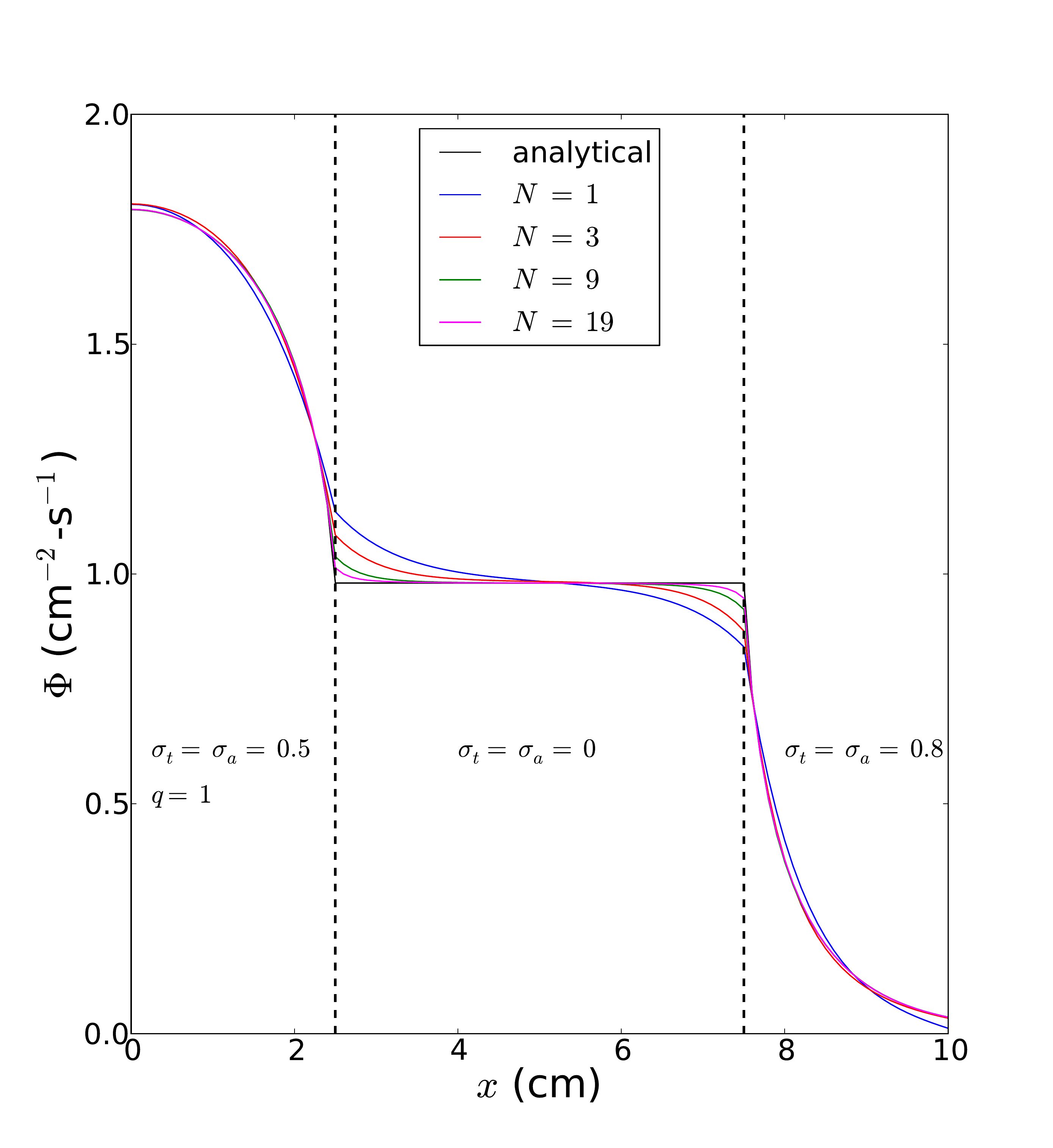}
		\subcaption{SAAF--CLS--\PN}
		\label{fig:1Dpb_CLS}
	\end{subfigure}
	\caption{Comparison of the scalar flux as a function of $x$ with and without the conservative fix described by Eq.\,\eqref{eq:conservativeFix} for different values of $N$.}
	\label{fig:1Dproblem}
\end{figure}

\begin{figure}
	\centering
	%\hspace*{-2cm}
	\includegraphics[width=0.8\textwidth]{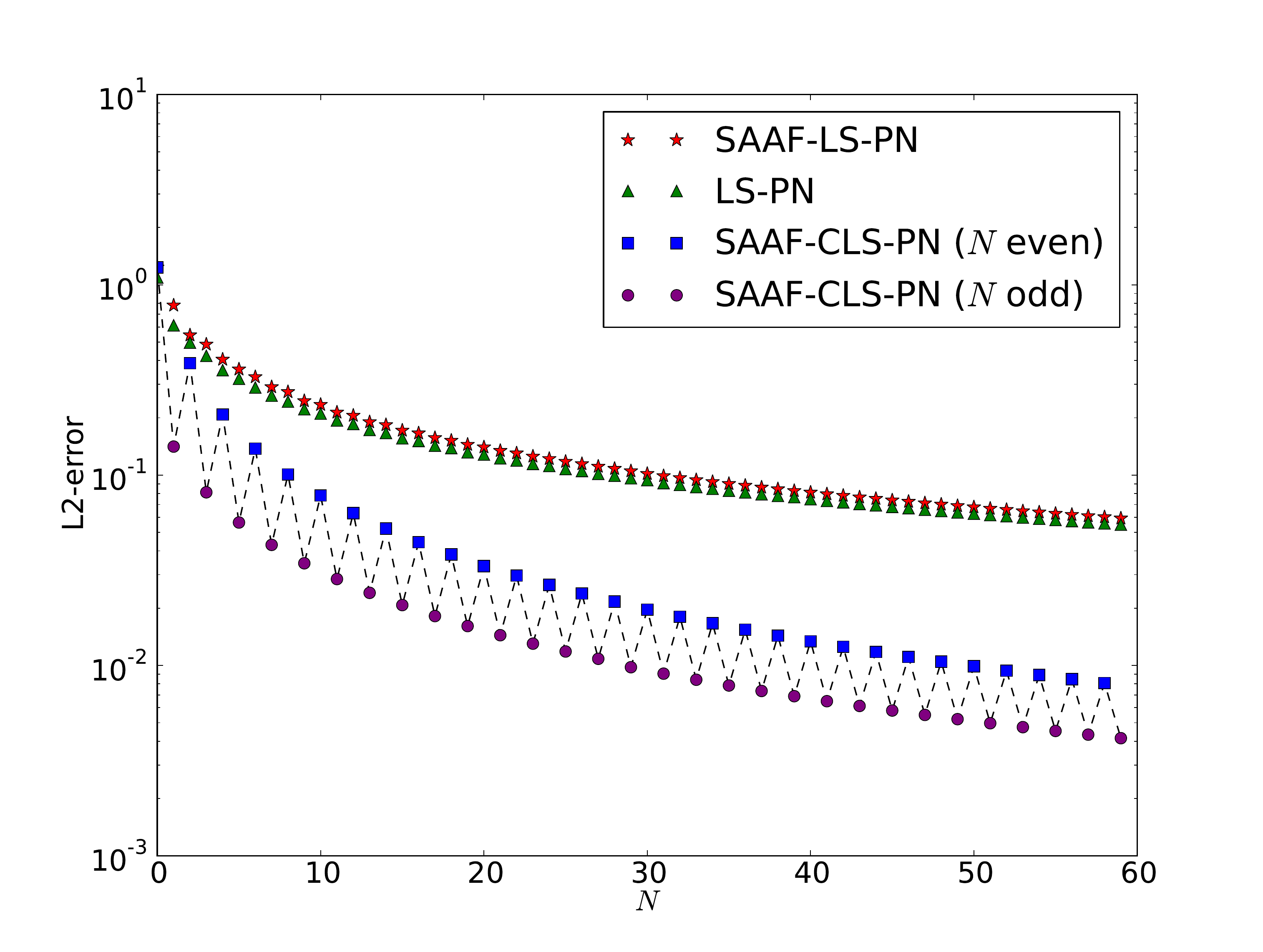}
	\caption{Comparison of the L2-error (in cm$^{-3/2}$--s$^{-1}$) of the scalar flux $\Phi$ for different discretizations. In particular, the SAAF--LS--\PN\,method (i.e.\,without the conservative fix in the void region) is comparable to the LS--\PN\,method. The SAAF--CLS--\PN\,method does much better, especially for odd values of $N$.}
	\label{fig:1Dproblem_error}
\end{figure}

In this section, we consider a scattering- and fission-free domain ($H=0$) composed of three distinct uniform regions, defined respectively for $0\leq x \leq \delta$, $\delta\leq x \leq 3\delta$ and $3\delta\leq x \leq 4\delta$. 
In the first one, $S\equiv q/w$ and $\sigma_t \equiv \sigma_{a,1}$; the second one is a pure void; in the third one, $S= 0$ and $\sigma_t \equiv \sigma_{a,2}$. 
The boundaries at $x=0$ and $x=4\delta$ respectively are reflecting and vacuum. 
The analytical scalar flux is given by:
\begin{equation}
\Phi(x) = \dfrac{q}{w\sigma_{a,1}}
\begin{cases}
\left(2-E_2\left(\sigma_{a,1}(\delta-x)\right)-E_2\left(\sigma_{a,1} (\delta+x)\right)\right) \quad&,\quad 0\leq x < \delta, \\
\left(1-E_2\left(2\sigma_{a,1}\delta\right)\right) \quad&,\quad \delta\leq x < 3\delta, \\
\left(E_2\left(\sigma_{a,2}(x-3\delta)\right) -E_2\left(2\sigma_{a,1}\delta + \sigma_{a,2}(x-3\delta)\right)\right) \quad&,\quad 3\delta\leq x < 4\delta, \\
\end{cases} 
\label{analyticalPhi}
\end{equation}
where $E_2$ represents the following exponential integral:
\begin{equation}
E_2\left(x\right) = \int_{1}^{\infty} \dfrac{\exp(-x z)}{z^2} \drm z.
\end{equation}
In practice, we choose $q=1$ cm$^{-3}$--s$^{-1}$, $\delta = 2.5$ cm, $\sigma_{a,1}$ = 0.5 cm$^{-1}$ and $\sigma_{a,2}$ = 0.8 cm$^{-1}$. 
A total of 4096 spatial cells is chosen. Fig.\,\ref{fig:1Dpb_LS} highlights why a conservative fix of the LS formulation is necessary (see Eq.\,\eqref{eq:conservativeFix}). 
Without it, the solution in void is clearly inaccurate and the convergence with $N$ is very slow. 
In particular, we mentioned that the LS formulation is globally conservative if and only if $(1 , \vo\cdot\grad\Psi )_{\D_0} = 0$ (see Eq.\,\eqref{eq:LSconservartion}), which is clearly not the case in the void region. 
Fig.\,\ref{fig:1Dpb_CLS} qualitatively shows the improvement in the results when using the conservative fix. 
Although we still have $\vo\cdot\grad\Psi \neq 0$ in the void region, the global conservation therein is maintained and the difference with the analytical solution appears to be greatly reduced. 
Fig.\,\ref{fig:1Dproblem_error} quantifies the L2-error with the analytical solution. 
Noteworthy is the fact that using the hybrid SAAF--LS method (without the fix) does not even outperform the plain LS method. 
However the SAAF--CLS method clearly does, as a SAAF--CLS--$P_{5}$ calculation gives an error almost identical to the LS--$P_{59}$ solution. 
Another interesting feature is that the parity of $N$ matters, odd values of $N$ giving better results for our hybrid method.

Table \ref{tab::VTvsCLS} compares the SAAF--CLS--\PN~and SAAF--VT--\PN~methods on that same problem. While the L2-errors of the numerical solution for a given $N$ is close for both methods, the number of GMRES iterations needed to converge grows very quickly for the latter one. This exhibits the conditioning problems it suffers from and makes it impractical for more complicated problems. \revision{The same quantities for the LS--\PN~method emphasizing once again that the L2--error is much higher than the other two methods. As for the number of iterations, it is even slightly higher than for the SAAF--CLS--\PN~for the values of $N$ shown. It turns out that for large values of $N$ (e.g.~$N\geq 19$), the number of iterations becomes lower for the LS--\PN~method.}

%\begin{figure}
	%\centering
%	\begin{subfigure}[b]{0.44\textwidth}
%		%\hspace*{-2cm}
%		\includegraphics[width=\textwidth]{GMRESiterations.eps}
%		\subcaption{Comparison of the number of iterations.}
%		\label{fig:GMRESiterations}
%	\end{subfigure}%
	%\quad %add desired spacing between images, e. g. ~, \quad, \qquad, \hfill etc.
	%(or a blank line to force the subfigure onto a new line)
%	\begin{subfigure}[b]{0.55\textwidth}
		\begin{table}[H]
			\begin{center}
				\begin{tabular}{|c||c|c|c||c|c|c|}
					\hline
					& \multicolumn{3}{|c||}{L2-error} & \multicolumn{3}{|c|}{Iteration count } \\
					\hline
                    $N$ & SAAF--CLS--\PN &  SAAF--VT--\PN &  $\;\;$ LS--\PN$\;\;$ & SAAF--CLS--\PN &  SAAF--VT--\PN & $\;\;$ LS--\PN$\;\;$ \\\hline
                    0	&	1.24E-0	&	1.52E-0	&	1.09E-0	&	9	&	7	&	5	\\
                    1	&	1.41E-1	&	1.64E-1	&	6.12E-1	&	35	&	801	&	20	\\
                    2	&	3.87E-1	&	5.96E-1	&	4.97E-1	&	50	&	1211	&	59	\\
                    3	&	8.12E-2	&	6.58E-2	&	4.24E-1	&	70	&	2765	&	79	\\
                    4	&	2.08E-1	&	3.68E-1	&	3.56E-1	&	84	&	4617	&	111	\\
                    5	&	5.64E-2	&	3.77E-2	&	3.21E-1	&	110	&	8120	&	118	\\
					\hline
				\end{tabular}
				\caption{\label{tab::VTvsCLS} Comparison of the SAAF--CLS--\PN, SAAF--VT--\PN~\revision{and LS--\PN~}methods in terms of the L2-error (in cm$^{-3/2}$--s$^{-1}$) and the number of GMRES linear iterations. Although the L2-error is fairly comparable for a given $N$ for the former two, the number of iterations rapidly becomes intractable for SAAF--VT--\PN. \revision{The number of iterations for the LS--\PN~method becomes lower than that of the SAAF--CLS--\PN~for large values of $N$ (e.g.~$N\geq19$).}}
			\end{center}
		\end{table}
%	\end{subfigure}
%	\caption{Comparison of the SAAF--VT--\PN~and SAAF--VT--\PN~methods in terms of L2-error and GMRES iterations.}
%	\label{fig:VTvsCLS}
%\end{figure}

\subsection{Modified Reed's Problem}
\label{sec:ReedsProblem}

\revision{In this section, we wish to study the impact of the scaling factor $c$ on the numerical solution. In particular, we consider the situation where $\sigt$ takes very different values on $\Gamma$, in which case the choice of $c$ is not obvious (see Section \ref{sec:valueOfc}).}

\revision{Specifically, we look at a famous test problem, known as the Reed's problem \cite{Reed1971} which has significant discontinuities between the different regions of the problem. To accentuate the discontinuity of $\sigt$ on $\Gamma$, we slightly modify the problem by reordering the spatial regions. Table \ref{table:reedSpecifications} summarizes the material properties of the modified problem, defined for $0\leq x\leq 8$ cm, with reflecting and vacuum boundary conditions respectively imposed at $x=0$ and at $x=8$ cm. We also choose 4096 cells, which makes the spatial error negligible for the values of $N$ considered in this section.}

\begin{table}[ht!]
	\centering
	\begin{tabular}{| c | c | c | c | c | c |}
		\hline
		& Region 1 & Region 2 & Region 3 & Region 4 & Region 5  \\[2ex] \hline
		$q$ & 100 & 0 & 0 & 0 & 1 \\
		$\sigt$ & 100 & 0 & 1 & 5 & 1 \\
		$\sigma_s$ & 0 & 0 & 0.9 & 0 & 0.9 \\[1ex]
		\hline
		Domain & $0\leq x < 2$ & $2\leq x < 4$ & $4\leq x < 6$ & $6\leq x < 7$ & $7\leq x \leq 8$ \\
		\hline
	\end{tabular}
	\captionof{table}{Material properties for the modified Reed's problem: value for the angular--integrated volumetric source $q$ in cm$^{-3}$--s$^{-1}$ ($S = q/w$) and the total and scattering cross-sections (in cm$^{-1}$) in each region. }
	\label{table:reedSpecifications}
\end{table}

\begin{figure}[ht!]
	\centering
	\begin{subfigure}[b]{0.55\textwidth}
		\hspace*{-0.5cm}
		\includegraphics[width=\textwidth, height=1\textwidth]{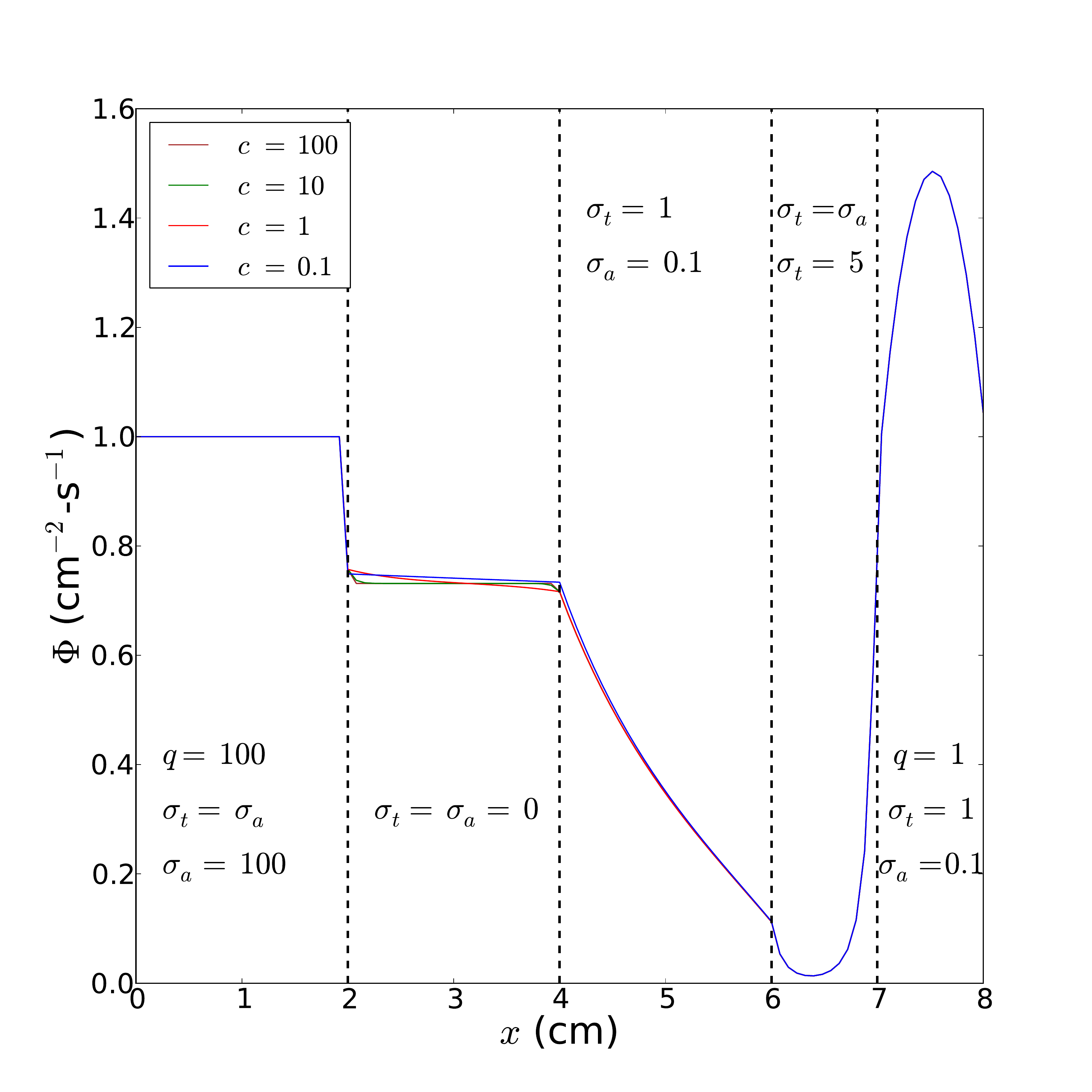}
		\subcaption{$P_3$}
		\label{fig:scattP3}
	\end{subfigure}%
	%\quad %add desired spacing between images, e. g. ~, \quad, \qquad, \hfill etc.
	%(or a blank line to force the subfigure onto a new line)
	\begin{subfigure}[b]{0.55\textwidth}
		\hspace*{-1cm}
		\includegraphics[width=\textwidth, height=1\textwidth]{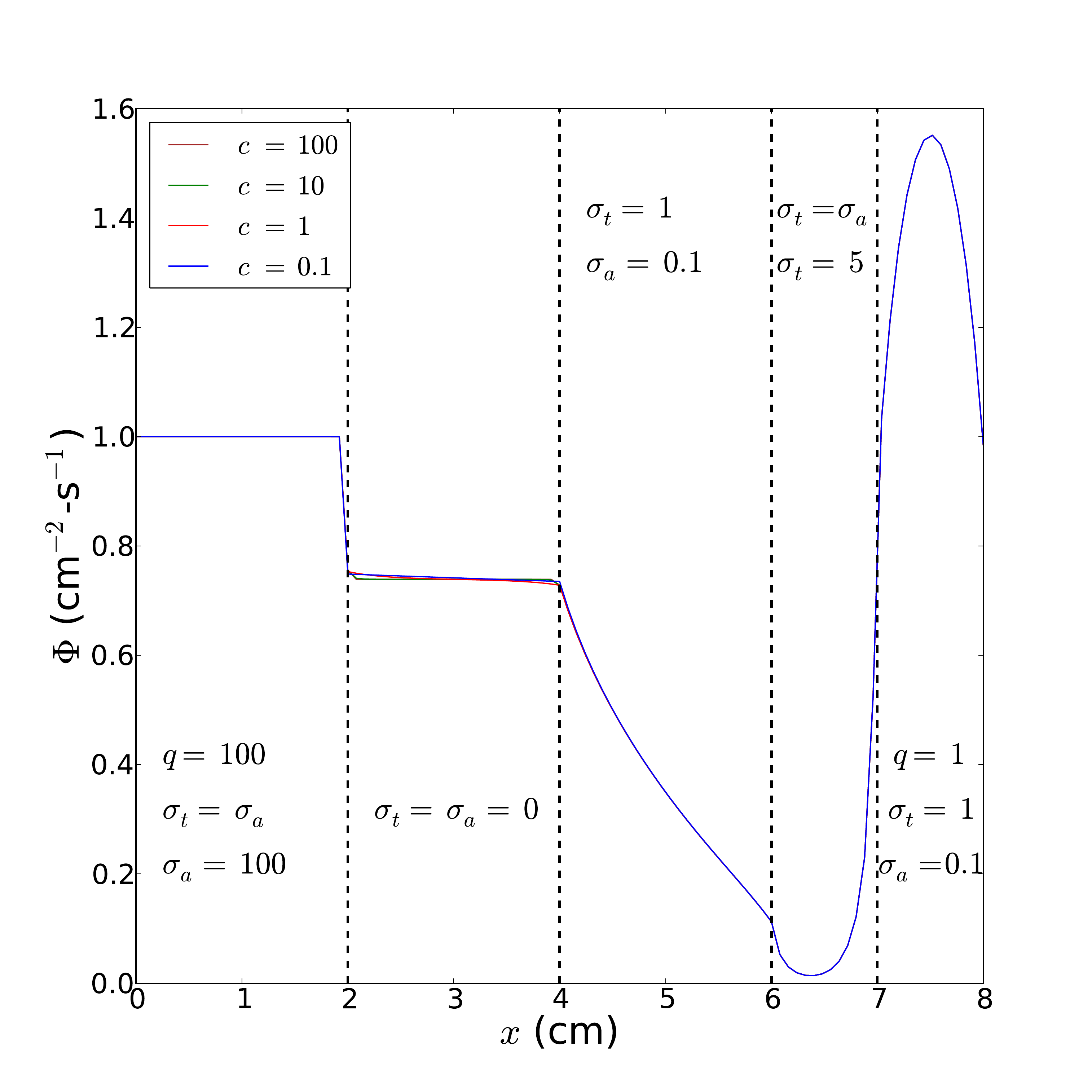}
		\subcaption{$P_7$}
		\label{fig:scattP7}
	\end{subfigure}               
	\caption{SAAF--CLS--\PN~method for different values of $c$ (in cm$^{-1}$) on the modified Reed's problem.}
	\label{fig:ReedsProblem}
\end{figure}

\revision{Fig.~\ref{fig:ReedsProblem} shows the results for the SAAF--CLS--\PN~method for different values of $c$. Fig.~\ref{fig:scattP3} indicates that it has a small impact on the $P_3$ numerical solution and that the difference is mostly limited to the void region, the solution elsewhere being virtually identical. Furthermore, the discrepancy is reduced as $N$ is increased, as the $P_7$ solution given on Fig.~\ref{fig:scattP7} shows.}

\revision{In conclusion, it indeed appears that the value of $c$ has little influence on the numerical solution\footnote{It can however have an impact on the conditioning of the system.} outside of the void region, which implies that the reaction rates are minimally affected. Moreover, this impact is further reduced as $N$ is increased.}

\subsection{Multigroup $k$-eigenvalue problem}
\label{sec:eigenvalue}

\begin{figure}[ht!]
	\begin{center}
		\includegraphics[width=0.65\textwidth]{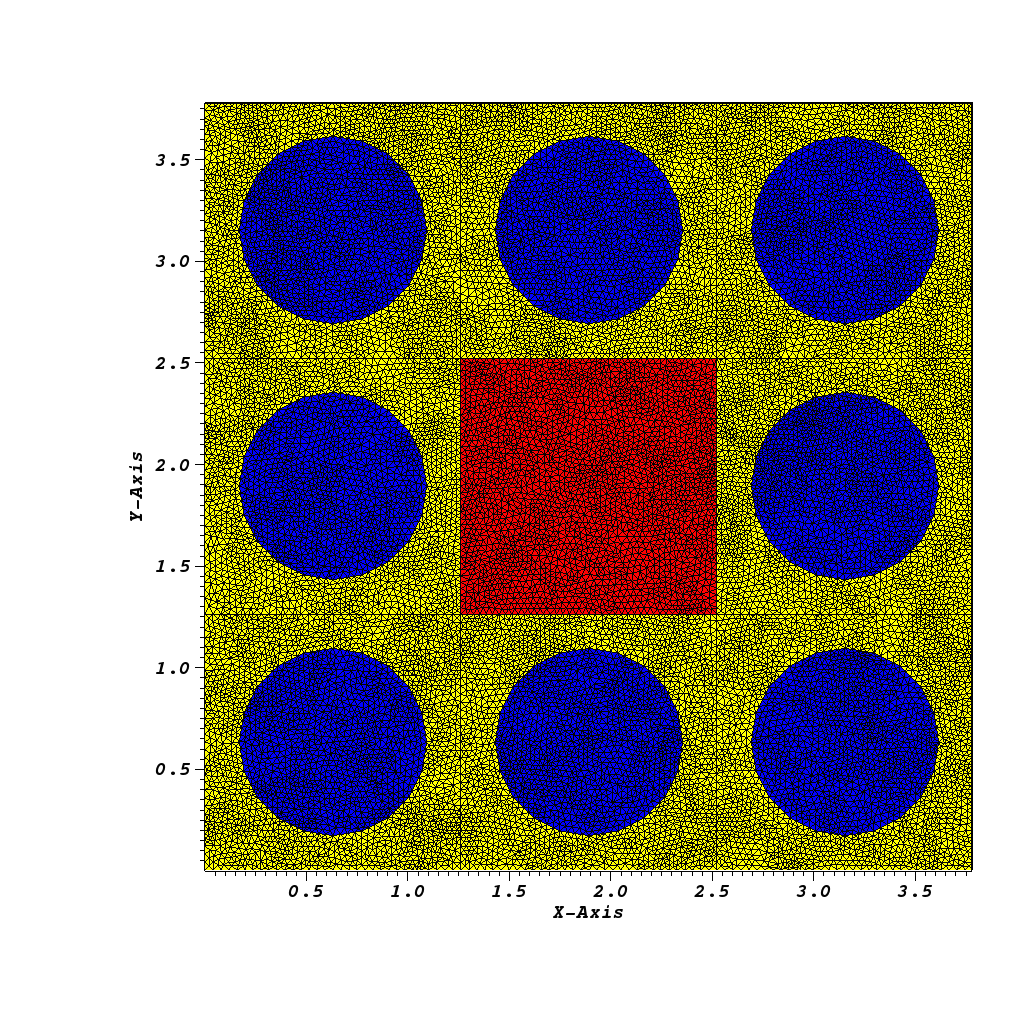}
		\caption[]{\label{fig:pinCellTestProblem}Geometry of a 3x3 pin cell test problem. The regions in blue, yellow and red respectively correspond to the fuel, moderator and void. The former two use the cross-sections of the C5G7 benchmark \cite{c5g7}. The latter is in practice chosen such that $\sigma_t = 10^{-10}$ cm$^{-1}$. The fuel boundary is approximated with a 20-side polygon. The meshes with a refinement of 0, 1, 2 (shown) and 3 have respectively 1116, 4829, 21090 and 92912 nodes. Besides, they respectively have 2134, 9455, 41776 and 184962 elements and 3249, 14283, 62865 and 277873 sides.}
	\end{center}
\end{figure}

We consider a test problem described in Fig.\,\eqref{fig:pinCellTestProblem} and already studied in \cite{LaboureWang2016} consisting of 8 pin cells surrounding a void region. 
The interest is to compare our methods on a multigroup heterogeneous test problem involving a void region.
Each of the 9 square subdomains is 1.2598 cm in length, assembled in a 3x3 configuration. 
The total size of the problem is therefore 3.7794 cm $\times$ 3.7794 cm. 
The subdomain in the center of the problem is void. The 8 others contain a pin of radius 0.45720 cm. 
Each pin boundary is approximated by a 20-side polygon.
The material properties of the fuel and moderator (shown in blue and yellow on the figure, respectively) are chosen to be identical to the "UO$_2$ Fuel-Clad mix" and "Moderator" materials from the C5G7 benchmark \cite{c5g7}. 
This problem is assumed to be infinite along the $z$ direction and therefore only depends on $x$ and $y$. 
The same problem\footnote{For convenience however, the pin boundaries for the MCNP calculation are circles.} was run using MCNP5 \cite{MCNP5} with 125 cycles of 10$^6$ particles (the first 25 cycles being discarded). 
The reference eigenvalue was estimated to be $\tilde{k}_{\text{eff}}=$ 1.34745 with a standard deviation of 5 pcm. 

Table \ref{tab::physorPN} shows the error with respect to $\tilde{k}_{\text{eff}}$ for the LS--\PN\,and the SAAF--CLS--\PN\,methods. 
The former, lacking global conservation, is extremely slow to converge as the number of elements and angular moments is increased \cite{LaboureWang2016}. 
The latter gives much better results as any solution with $N\geq 3$ yields an error in $k_{\text{eff}}$ smaller than the LS--$P_{39}$ solution, which has over 5.3$\times 10^8$ unknowns\footnote{The number of unknowns is the product of the number of nodes $n$, moments $L=(N+1)(N+2)/2$ and energy groups $G=7$.}.
In particular, the most refined calculations are within a few standard deviations. 
%It is expected that the most refined meshes would give the best results, if the value of $N$ were to be increased even further. 
%However, the practical number of MOOSE kernels \cite{MOOSE} and thus of $N$ are limited by the memory each individual processor is allocated.

Table \ref{tab::physorSN} compares the same quantity for the LS--\SN, SAAF--CLS--\SN\,and SAAF--VT--\SN\,methods as a function of the number of polar and azimuthal angles per quadrant, noted $N_p$ and $N_a$ respectively. 
The quadrature rule used is the Bickley3-Optimized\footnote{The total number of angles per quadrant is then $N_a N_p$.} because it appeared to converge faster than the Level-Symmetric quadrature rule. 
Once again, the LS method does a lot worse than the other two globally conservative methods. In particular, the spatial error dominates the calculations shown if $N_p\geq 2$.
It is interesting to note that SAAF--CLS--\SN\,and SAAF--VT--\SN\,give very similar results, which is expected since the two variational formulations are so close.

\begin{table}[H]
	\begin{center}
		\begin{tabular}{|c|cccc|cccc|}
			\hline
			& \multicolumn{4}{|c|}{SAAF--CLS--\PN} & \multicolumn{4}{|c|}{LS--\PN} \\
			\hline
			$\; N \;$ & Ref = 0 & Ref = 1 & Ref = 2 & Ref = 3 & Ref = 0 & Ref = 1 & Ref = 2 & Ref = 3 \\
			\hline
			1	&	946	&	903	&	894	&	892	&	56391	&	52543	&	51219	&	50676	\\
			3	&	312	&	227	&	208	&	238	&	24370	&	20159	&	19083	&	18771	\\
			5	&	111	&	14	&	-8	&	-14	&	15091	&	10382	&	9208	&	8901	\\
			7	&	50	&	-48	&	-72	&	-78	&	12107	&	7063	&	5755	&	5413	\\
			9	&	38	&	-59	&	-83	&	-89	&	10705	&	5407	&	3974	&	3586	\\
			19	&	66	&	-23	&	-42	&	-47	&	9431	&	3625	&	1845	&	1273	\\
			29	&	79	&	-6	&	-23	&	-25	&	9163	&	3223	&	1352	&	722	\\
			39	&	85	&	1	&	-16	&	-18	&	9087	&	3110	&	1210	&	561	\\
			\hline
		\end{tabular}
		\caption{\label{tab::physorPN} Error $k_{\text{eff}}-\tilde{k}_{\text{eff}}$ (in pcm). "Ref" designates the mesh refinement level. The standard deviation on $\tilde{k}_{\text{eff}}$ is 5 pcm.}
	\end{center}
\end{table}

\begin{table}[H]
	\begin{center}
		\begin{tabular}{|c|ccc|ccc|ccc|}
			\hline
			 & \multicolumn{3}{|c|}{SAAF--CLS--\SN} & \multicolumn{3}{|c|}{SAAF--VT--\SN} & \multicolumn{3}{|c|}{LS--\SN} \\
			\hline
			$\;(N_p,N_a) \;$ & Ref = 0 & Ref = 1 & Ref = 2 & Ref = 0 & Ref = 1 & Ref = 2 & Ref = 0 & Ref = 1 & Ref = 2  \\
			\hline
			(1,\,12) &	350	&	276	&	260	&	337	&	276	&	271	& 8570	&	3045	&	1284	\\
			(1,\,24)&	357	&	291	&	278	&	355	&	289	&	278	& 8561	&	3079	&	1316	\\
			(1,\,48) &	358	&	294	&	282	&	359	&	296	&	287	& 8530	&	3068	&	1320	\\
			(1,\,96) &	358	&	294	&	283	&	359	&	296	&	288	& 8521	&	3062& 1318		\\\hline
			(2,\,12)	&	91	&	2	&	-17	&	67	&	-2	&	-8	&	9042	&	2955	&	1045	\\
			(2,\,24)	&	98	&	15	&	-2	&	83	&	9	&	-2	&	9033	&	2988	&	1075	\\
			(2,\,48)	&	98	&	18	&	3	&	87	&	16	&	6	&	9004	&	2978	&	1080	\\
			(2,\,96)	&	98	&	18	&	3	&	87	&	17	&	7	&	8995	&	2972	&	1077	\\\hline
			(3,\,12)	&	81	&	-11	&	-31	&	54	&	-16	&	-22	&	9121	&	3008	&	1056	\\
			(3,\,24)	&	87	&	3	&	-16	&	71	&	-4	&	-16	&	9112	&	3042	&	1087	\\
			(3,\,48)	&	88	&	6	&	-11	&	75	&	3	&	-8	&	9083	&	3032	&	1091	\\
			(3,\,96)	&	88	&	6	&	-11	&	75	&	3	&	-7	&	9074	&	3025	&	1089	\\
			\hline
		\end{tabular}
		\caption{\label{tab::physorSN} Error $k_{\text{eff}}-\tilde{k}_{\text{eff}}$ (in pcm). "Ref" designates the mesh refinement level. The standard deviation on $\tilde{k}_{\text{eff}}$ is 5 pcm.}
	\end{center}
\end{table}

%\begin{table}[H]
%	\begin{center}
%		\begin{tabular}{|ccc|c|}
%			\hline
%			Number of triangles & $\qquad N_p\qquad$ & $\qquad N_a\qquad$ & $\qquad k_\text{eff}- \tilde{k}_{\text{eff}}\qquad$  \\
%			\hline
%			4454 & 2 & 6 & 27 \\
%			22024 & 2 & 6 & 24 \\
%			22024 & 4 & 12 & -4 \\
%			22024 & 8 & 24 & -17 \\
%			44767 & 8 & 24 & -17 \\
%			\hline
%		\end{tabular}
%		\caption{\label{tab::PDT}Value for $k_\text{eff}-\tilde{k}_{\text{eff}}$ (in pcm) with PDT (Bi-Linear Discontinuous \SN-DFEM code). The standard deviation on $\tilde{k}_{\text{eff}}$ is 5 pcm.}
%	\end{center}
%\end{table}

\subsection{Dog leg void duct problem}
\label{sec:kobayashi}

\begin{figure}[H]
	\centering
	%\hspace*{-2cm}
	\includegraphics[width=0.4\textwidth]{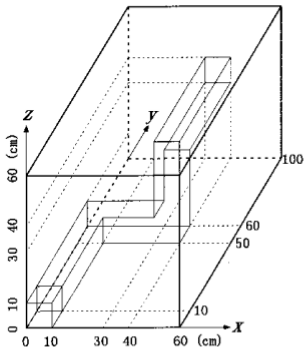}
	\caption{Geometry of the dog leg void duct problem (figure taken from \cite{Kobayashi2001}) }
	\label{fig:kobayashi}
\end{figure}

In this section, we consider the third benchmark problem introduced by Kobayashi et al in \cite{Kobayashi2001}, also called the dog leg void duct problem. Here, we only show results for the pure absorber problem.
The rectangular spatial domain is defined for $0\leq x,z\leq 60$ cm and $0\leq y\leq 100$ cm. 
The geometry of the problem is shown in Fig.\,\ref{fig:kobayashi} and consists of three uniform materials. First, a source region for $\max(x,y,z)\leq 10$ cm with a volumetric source $S=1$ cm$^{-3}$-s$^{-1}$ and $\sigt=\sigma_a = 0.1$ cm$^{-1}$. 
Second, a near-void region ($\sigt=\sigma_a = 10^{-4}$ cm$^{-1}$) for $0\leq x,z\leq 10$ cm and $10\leq y\leq60$ cm; $10\leq x\leq40$ cm, $50\leq y\leq60$ cm and $0\leq z\leq10$ cm;  $30\leq x\leq40$ cm, $50\leq y\leq60$ cm and $10\leq z\leq40$ cm; $30\leq x\leq40$ cm, $60\leq y\leq100$ cm and $30\leq z\leq40$ cm. 
Third, a shield region defined everywhere else by $\sigt=\sigma_a = 0.1$ cm$^{-1}$. 
Reflecting boundary conditions are imposed at $x =0$, $y =0$ and $z =0$; vacuum boundary conditions are used at $x =60$ cm, $y =100$ cm and $z =60$ cm. 
The coarsest mesh used has $6\times 10\times 6$ cube elements and is referred to as the "ref = 0" mesh. %Increasing the level of mesh refinement by one essentially multiplies the number of elements by eight.  

The interest of this problem lies not only in comparing the different methods on a widely-studied 3-D benchmark problem but also in testing our new method in near-void regions, while the previous two problems only had pure void regions.
 
In this section, all the \SN\,simulations use the Level-Symmetric angular quadrature rule and the total number of angles is then given by $N(N+2)$, as the solution depends on all three spatial variables. 
As a comparison, the total number of moments for a \PN\,simulation in that case is $(N+1)^2$, which implies that, for a given $N$, the number of angular unknowns only differs by one between a \PN\,and a \SN\,calculation.

\begin{figure}
	%\centering
	\begin{subfigure}[b]{0.49\textwidth}
		%\hspace*{-2cm}
		\includegraphics[width=\textwidth]{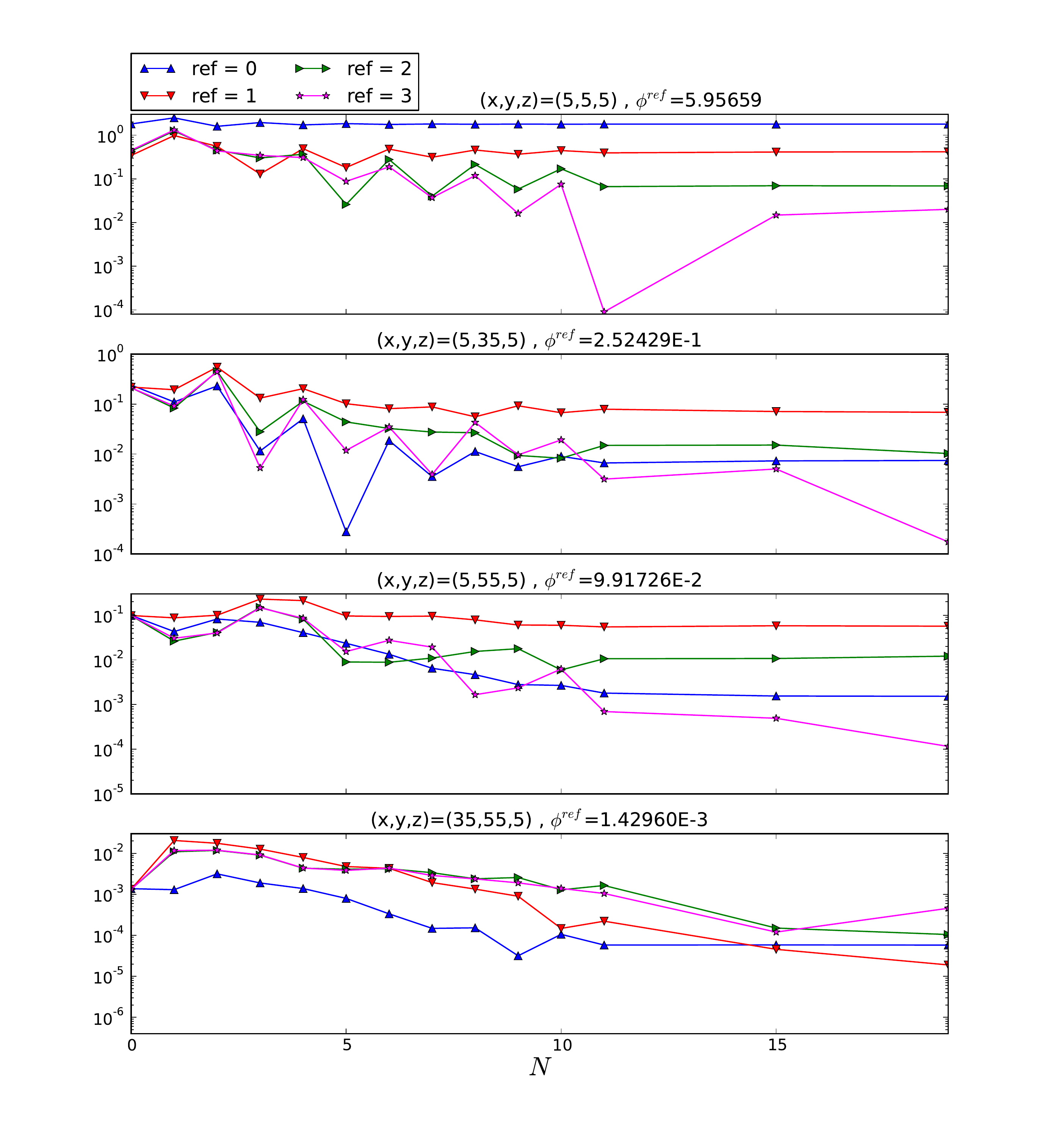}
		\subcaption{SAAF--CLS--\PN}
		\label{fig:kobayashi_SAAFCLSPN}
	\end{subfigure}%
	\begin{subfigure}[b]{0.49\textwidth}
		%\hspace*{-2cm}
		\includegraphics[width=\textwidth]{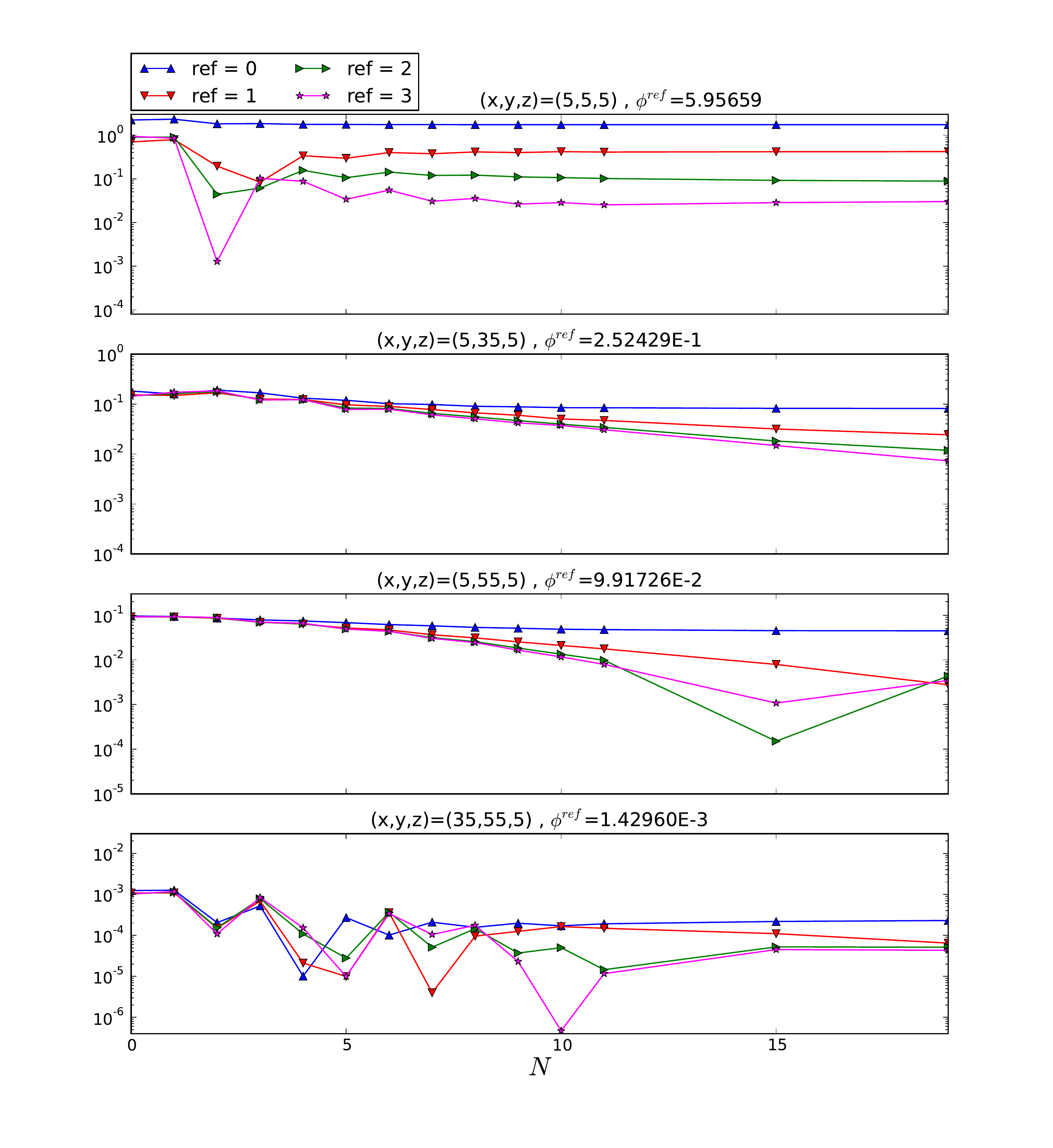}
		\subcaption{LS--\PN}
		\label{fig:kobayashi_LSPN}
	\end{subfigure}
	\vspace{0cm}\\
	\begin{subfigure}[b]{0.49\textwidth}
		%\hspace*{-2cm}
		\includegraphics[width=\textwidth]{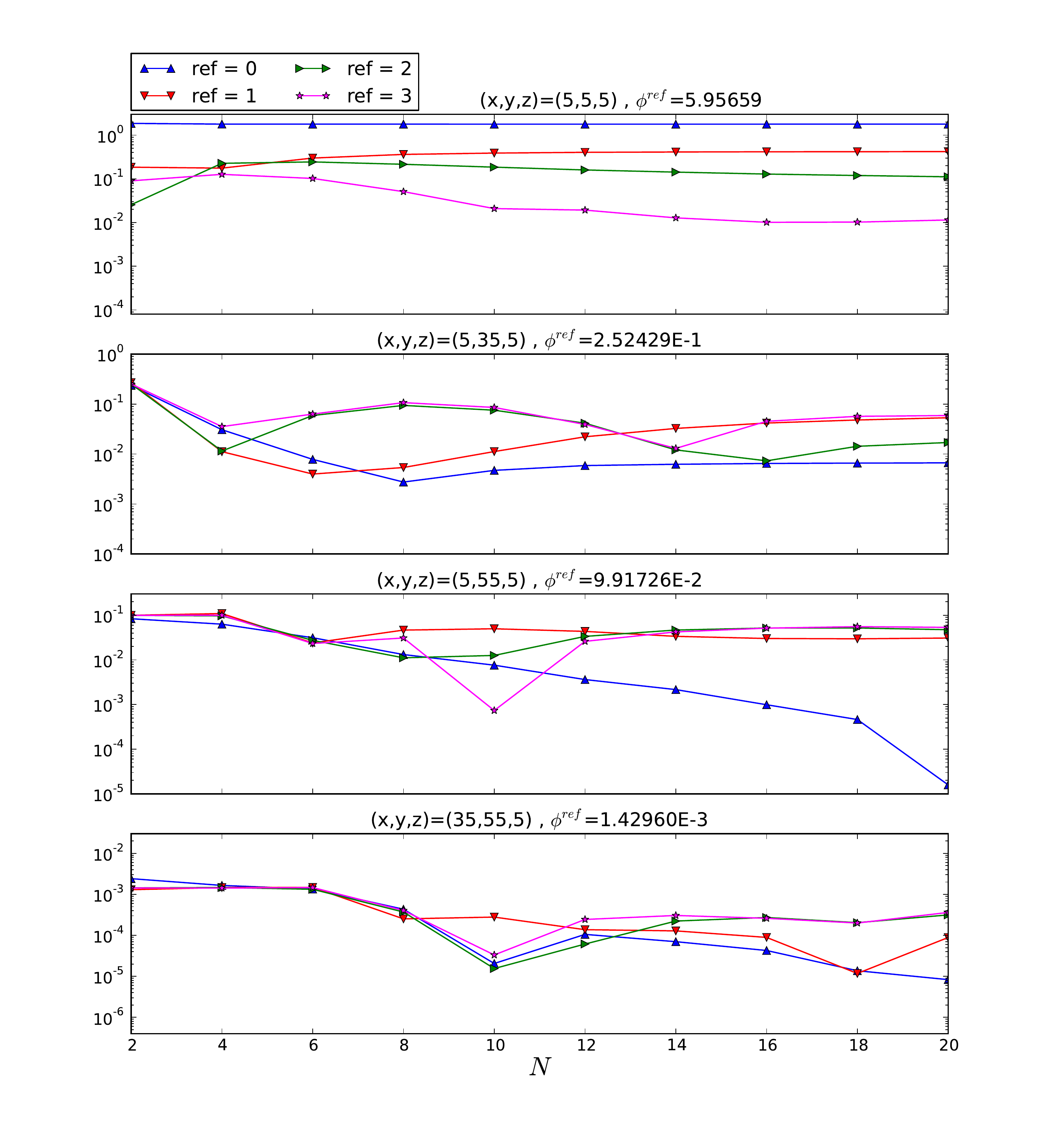}
		\subcaption{SAAF--CLS--\SN}
		\label{fig:kobayashi_SAAFCLSSN}
	\end{subfigure}%
	%\quad %add desired spacing between images, e. g. ~, \quad, \qquad, \hfill etc.
	%(or a blank line to force the subfigure onto a new line)
	\begin{subfigure}[b]{0.49\textwidth}
		%\hspace*{-2cm}
		\includegraphics[width=\textwidth]{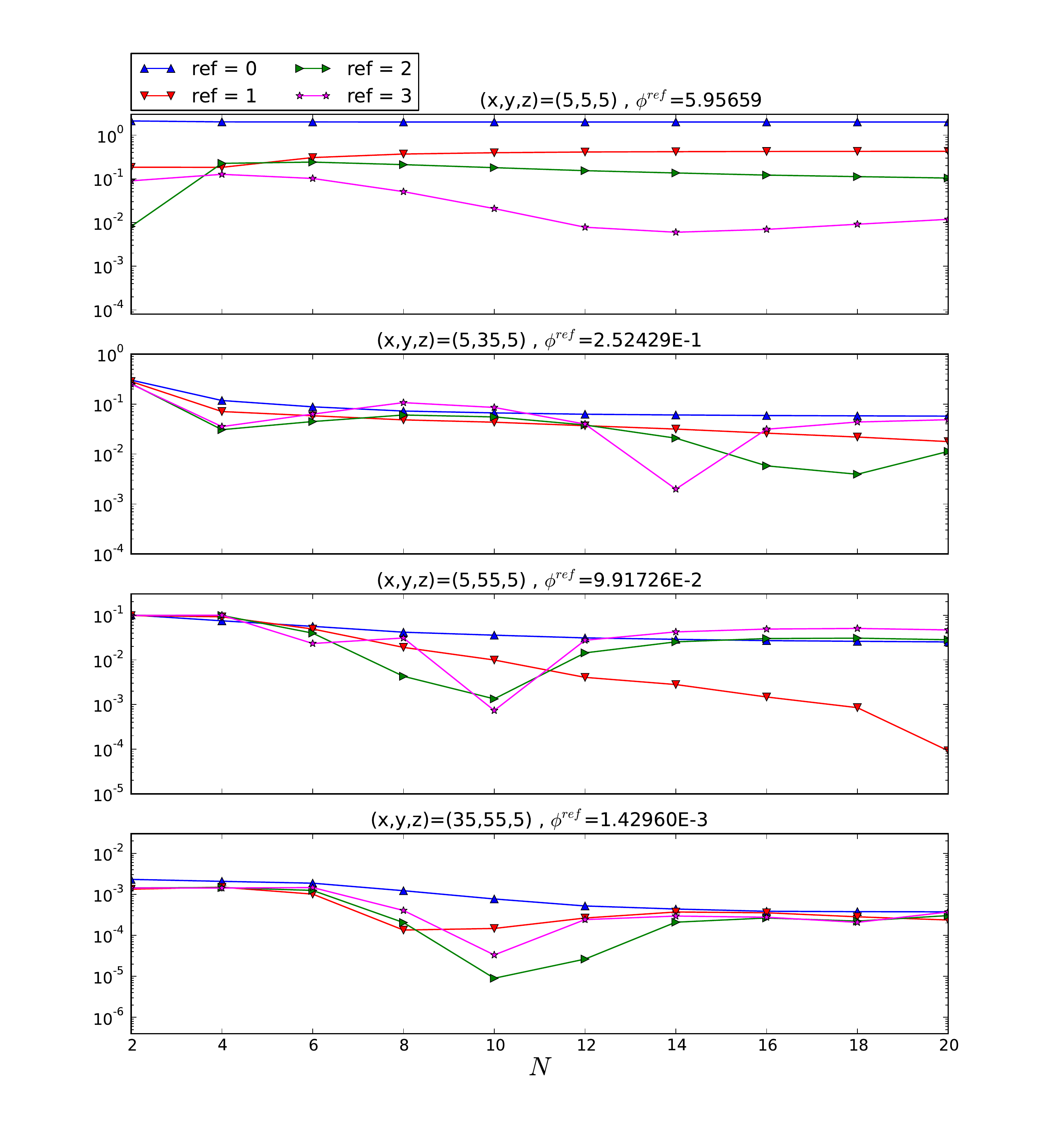}
		\subcaption{SAAF--VT--\SN}
		\label{fig:kobayashi_SAAFVTSN}
	\end{subfigure}
	\caption{Error in the scalar flux (in cm$^{-2}$--s$^{-1}$) at 4 different spatial points as a function of $N$ and of the level of mesh refinement for various methods. On the top of each graph, the reference value for $\Phi$ at the corresponding spatial point is indicated. For a given spatial point, the $y$-axis range is identical for all methods.}
	\label{fig:koba}
\end{figure}

In \cite{Kobayashi2001}, the semi-analytical scalar flux was given at different points of the domain. 
In Fig.\,\ref{fig:kobayashi_SAAFCLSPN}, we compute the SAAF--CLS-\PN\,error at $(x,y,z)=(5,5,5)$, $(5,35,5)$, $(5,55,5)$ and $(35,55,5)$ for different values of $N$ and of the level of mesh refinement. The first point is in the source region whereas the last three are in the near-void region.
It appears that the error indeed decreases as the simulation is refined in space and angle, although it becomes less apparent for the spatial point $(35,55,5)$, further away from the source. 
This is not surprising as the magnitude of the scalar flux rapidly decreases in the shield region. 
As we observed in Fig.\,\ref{fig:1Dproblem_error}, the error seems to be generally higher for even values of $N$. 

Fig.\,\ref{fig:kobayashi_LSPN} shows the results for the LS-\PN\,method which are very comparable to SAAF--CLS-\PN\,at the first and last spatial points but somewhat worse at the second and third point. The difference is not as significant as in Fig.\,\ref{fig:1Dproblem}, most likely because the near-void region is spatially much more limited than it was in Section \ref{sec:1Dproblem}, where the void region accounted for half of the spatial domain.

In Fig.\,\ref{fig:kobayashi_SAAFCLSSN}, the same quantities are shown for the SAAF--CLS-\SN\,method, with errors comparable to SAAF--CLS-\PN\,at the first and last spatial points but not as good at the second and third. 
It is noted however that the computational time tend to be much lower for the \SN\,method, in particular because the number of kernels needed to be assembled for the streaming terms is noticeably higher for \PN, due to numerous off-diagonal coupling terms.

Lastly, Fig.\,\ref{fig:kobayashi_SAAFVTSN} exhibits a behavior for the SAAF--VT-\SN\,method very close to that of SAAF--CLS-\SN, especially for a level of spatial refinement higher than 2.
This was expected as both variational formulations look very much alike (see Eqs.\,\eqref{SAAFCLS} and \eqref{eq:SAAFVT}). 
For instance, the error at the first spatial point for the most refined mesh approaches $10^{-2}$ cm$^{-2}$-s$^{-1}$ in both cases. 
However, it is interesting to point out that the Hypre BoomerAMG preconditioner \cite{hypre} does not seem to be efficient in the same ranges for both methods, most likely because of the on-diagonal contribution of the $(\vo\cdot\grad\Psi^*,c^{-1}\vo\cdot\grad\Psi)_{\D_0}$ term in the void region which does not vanish as the mesh is refined.

%------------------------------------------------------------------------------
%
%------------------------------------------------------------------------------
\section{Conclusion} 
\label{sec:conclusion}

In conclusion, we have derived a second-order method compatible with void and globally conservative. 
This is achieved by using LS terms in the void region with a non-symmetric correction to retrieve global conservation.
It is then combined with SAAF terms in the non-void regions with a scaling chosen such that the interface terms vanish, thereby maintaining global conservation. 
We have observed that this conservative fix is crucial to gain any benefit, compared to the plain LS method. 
Overall, this SAAF--CLS method has shown much improvement for problems for which global conservation is key, such as regions with large void regions or $k$-eigenvalue calculations. 
Particularly, we have obtained very satisfying results for both the \PN\,and \SN\,versions of our method on a multigroup $k$-eigenvalue problem with significant heterogeneity. 
These results have been in good agreement with a MCNP reference calculation but also have been very comparable to those obtained with the SAAF--VT--\SN\,method.

While the SAAF--CLS and SAAF--VT variational formulations look a lot alike, both sacrificing the symmetry of the bilinear form, our method presents the advantage of being compatible with both \PN\,and \SN\,angular discretizations, unlike the SAAF--VT method which has only shown success with \SN\,discretizations.
The reason is that the latter method tends to reduce to a first-order form in void regions, which results in a singular system following a \PN\,discretization for a steady-state calculation with CFEM.

Further, we have generalized the SAAF--CLS method to near-void regions and showed that global conservation could be preserved, providing a slightly different correction to the LS formulation. 
We have then tested this method on the dog leg void duct benchmark problem by Kobayashi et al \cite{Kobayashi2001}.

In the future, we wish to extend this method to time-dependent problems. 
Further work may also include deeper analysis studying if there is an optimal value for the constant $c$.

%In conclusion, we have derived boundary conditions for LS which are consistent to the SAAF weak formulation with void treatment \cite{ls-sn}, independently of the angular discretization. With such boundary conditions, LS and SAAF are then equivalent if the cross-section $\sigma_t$ is constant over the spatial domain and if the mesh is uniform. Our method gave the expected convergence behaviors in space and angle. %
%In particular, solving only for the even-parity component of the angular flux, when possible, decreases the spatial order of convergence by one. %
%However, being globally non-conservative, LS differs noticeably on heterogeneous problems from the SAAF results before convergence is achieved.

\section*{Acknowledgments}
We are very thankful to Pablo Vaquer for providing the MCNP reference solution for the multigroup $k$-eigenvalue problem in Section \ref{sec:eigenvalue}. 
%Likewise, we acknowledge the precious help of Tarek Ghaddar and Carolyn McGraw for running the same problem with the first-order \SN\,code PDT.

\bibliographystyle{elsarticle-num}
\bibliography{bibliography}
	
\end{document}